\documentclass[twocolumn,prl]{revtex4}
\usepackage{amsmath}
\usepackage{graphicx}
\usepackage{bm}
\usepackage{color}
\usepackage{amsfonts}
\usepackage{dcolumn}

\begin{document}
\title
{Fundamental differences between glassy dynamics in two and three dimensions}

\author{Elijah Flenner and Grzegorz Szamel}
\affiliation{Department of Chemistry, Colorado State University, Fort Collins, CO 80523}
\date{\today}
\begin{abstract}
The two-dimensional freezing transition is very different
from its three-dimensional counterpart. In contrast, the
glass transition is usually assumed to have similar characteristics in
two and three dimensions. Using computer simulations we show that 
glassy dynamics in supercooled two- and three-dimensional fluids are 
fundamentally different. Specifically, transient localization of particles upon
approaching the glass transition is absent in two dimensions, whereas it is
very pronounced in three dimensions. Moreover, the temperature dependence of the 
relaxation time of orientational correlations is decoupled from that of 
the translational relaxation time in two dimensions but
not in three dimensions. Lastly, the relationships between the 
characteristic size of dynamically heterogeneous regions and the relaxation time
are very different in two and three dimensions. 
These results strongly suggest that the glass transition 
in two dimensions is different than in three dimensions. 
\end{abstract}

\maketitle

\section{Introduction}
In two-dimensional (2D) solids, thermal fluctuations destroy crystalline 
order, displacement correlations increase logarithmically,
and density correlations decay according to power laws \cite{Strandburg1988,Mermin1968}. 
However, there can be long-range bond-orientational order in 2D \cite{Mermin1968}. 
The transition from the 2D fluid phase to the solid phase
can occur in two steps with an intermediate phase characterized by an exponential decay of the
density correlations and a power-law decay of the bond-orientational 
correlations \cite{Strandburg1988,Bernard2011}.
In contrast, in three-dimensional (3D) solids fluctuations do not destroy crystalline order \cite{Peierls1935}, 
and long-range translational and
rotational order emerge together at the freezing transition.  

Despite these differences between two- and three-dimensional ordered 
solids, the formation of an \textit{amorphous} solid upon supercooling a fluid, 
\textit{i.e.} the glass transition, 
is generally assumed to have similar characteristics in 2D and 3D \cite{Harrowell2006}. 
This assumption is reflected in
the trivial dimensional dependence of most glass transition theories \cite{Berthier2011}. 

We show that structural relaxation of supercooled fluids in two dimensions is different than in three dimensions. 
While we find the transient localization often associated with glassy dynamics in three dimensions, 
we do not find any transient localization in two dimensions if we simulated systems large enough to 
remove any finite size effects. 
Furthermore, the temperature dependence of the bond-orientational correlation time is decoupled from that 
of the translational relaxation time  
time in two dimensions, but these relaxation times have very similar temperature dependence in three dimensions.  
Along with these differences in structural relaxation, we also find that the characteristic size of regions of 
correlated mobility, dynamic heterogeneities, increases faster with the structural relaxation time in two 
dimensions than in three dimensions, and these regions are more ramified in two dimensions than three dimensions. 
Lastly, we show that the structural relaxation and heterogeneous dynamics depends on the underlying dynamics 
in two dimensions. 

\section{Results}

\begin{figure*}
\includegraphics[width=7.5in,bb=0.3in 4.5in 7.0in 7.4in,clip=True]{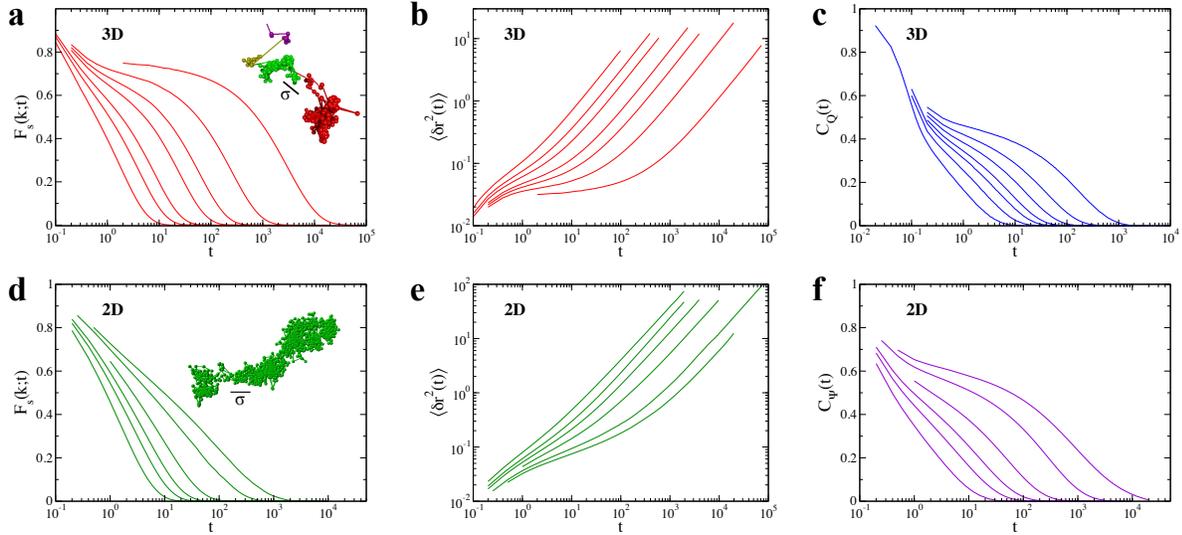}
\caption{\label{cage}\textbf{Structural relaxation in two and three dimensions.} 
{\bf a}, The self-intermediate scattering function $F_s(k;t)$ for the 3D glass-former for 
$T=1.0$, 0.8, 0.7, 0.6, 0.55, 0.5, and 0.45
listed from left to right.
The mode-coupling temperature $T_c^{3D} \approx 0.435$. The inset shows a trajectory plot of one small particle 
at $T=0.45$ where the color of the trajectory plot changes when the particle moves more than one large particle
diameter $\sigma$ (length of the black line) over a time of $0.1 \tau_\alpha$. 
{\bf b}, The mean square displacement $\left< \delta r^2(t) \right>$ 
for the 3D system showing the same temperatures as in ({\bf a}).
{\bf c}, The bond-orientational correlation function $C_Q(t)$ for the 3D glass-former 
showing the same temperatures as in ({\bf a}).
{\bf d}, The self-intermediate scattering function $F_s(k;t)$ 
for the 2D glass-former for $T=1.0$, 0.8, 0.7, 0.6, 0.5, and 0.45 listed from left to right.
The inset shows a trajectory plot of a small particle, and no sudden jumps are observed.
{\bf e}, The mean square displacement $\left< \delta r^2(t) \right>$ 
for the 2D system showing the same temperatures as in ({\bf d}).
{\bf f}, The bond-orientational correlation function $C_\Psi(t)$ for the 2D glass-former
showing the same temperatures as in ({\bf d}).
}
\end{figure*}

\begin{figure*}
\includegraphics[width=7.5in,bb=0.3in 5.75in 7.0in 7.4in,clip=True]{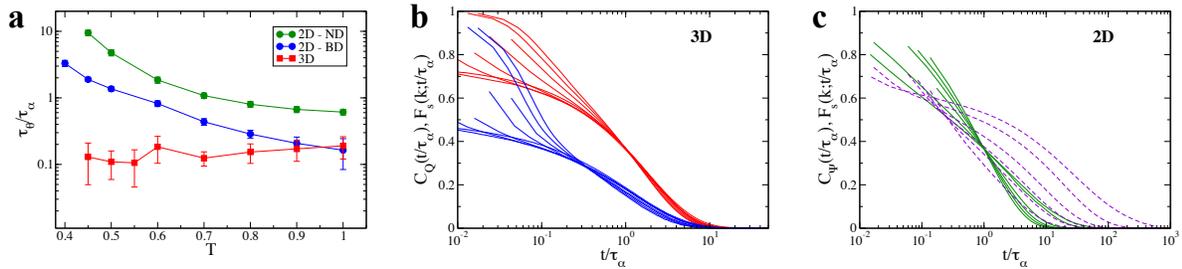}
\caption{\label{bond}\textbf{Bond angle and translational relaxation times.}
{\bf a}, The ratio of the relaxation time for the bond-orientational correlation functions
$\tau_\theta$ and the self-intermediate scattering function $\tau_\alpha$ for
the 2D system (circles) and the 3D system (squares). The ratio $\tau_\theta/\tau_\alpha$ for 
the 3D system is approximately constant and equal to 0.1-0.2 over the 
entire range of temperatures.  The green circles are for
Newtonian dynamics (ND) and the blue circles are results for Brownian dynamics (BD).
{\bf b}, The self-intermediate scattering function $F_s(k;t)$ (red solid lines)
and the bond-angle time correlation function $C_Q(t)$ (blue lines) 
rescaled by $\tau_\alpha$ for the 3D system.
{\bf c}, The self-intermediate scattering function $F_s(k;t)$ (green solid lines) and the bond-angle time
correlation function $C_\Psi(t)$ (violet dashed lines) rescaled by $\tau_\alpha$ for the 2D system.
 }
\end{figure*}

{\bf Structural relaxation.} 
To demonstrate the differences between
glassy dynamics in two and three dimensions, we 
focus on two closely related glass-forming fluids: 
the 3D 80:20 binary Lennard-Jones
system introduced by Kob and Andersen \cite{Kob1994}, and its 2D
variant that has the same interaction potentials but a 65:35 composition
to avoid crystallization \cite{Bruning2009}. To simulate the relaxation in these systems we used the standard
Newtonian dynamics \cite{AT}. We also simulated the 2D system using Brownian dynamics \cite{AT}
and we comment on the differences between these two dynamics. 
See Methods for the simulation details and the reduced units used to present the results. 
We examined three other 2D glass formers and
one additional 3D glass former (see Methods for details of the systems) and present  
results, which are qualitatively the same as the results for the KA system, in the supplemental material. 

In 3D, the dominant feature in the dynamics of deeply supercooled glass-forming fluids  
is transient localization of individual particles \cite{Berthier2011}, which is illustrated in the inset to 
Fig.~\ref{cage}a.
The transient localization results in characteristic plateaus of the
self-intermediate scattering function, 
$F_s(k;t) = N^{-1} \left< \sum_n e^{i \mathbf{k} \cdot [\mathbf{r}_n(t) - \mathbf{r}_n(0)]}\right>$ 
(k=7.2 in 3D and 6.28 in 2D), shown in
Fig.~\ref{cage}(a), and the mean-square displacement, 
$\left< \delta r^2(t) \right> = N^{-1} \left< \sum_n [\mathbf{r}_n(t) - \mathbf{r}_n(0)]^2 \right>$, shown in 
Fig.~\ref{cage}b. The plateaus extend to longer and longer times upon approaching the glass transition.
Similar plateaus are observed in the collective scattering function,
$F(k;t) = N^{-1} \left< \sum_{n,m} e^{i \mathbf{k} \cdot [\mathbf{r}_n(t) - \mathbf{r}_m(0)]}\right>$,
which describes relaxation of the density field (not shown), 
and in the correlation function quantifying bond-orientational 
correlations in 3D, $C_Q(t)$,  shown in Fig.~\ref{cage}c (see Methods for the definition of $C_Q$).
Qualitatively similar slowing down of the translational and
bond-orientational relaxation in 3D glass forming fluids is analogous to the simultaneous appearance of 
translational and rotational long range order in 3D crystalline solids.

The transient localization observed in 3D glassy dynamics is absent in 2D, 
as showed in the inset to 
Fig.~\ref{cage}d. 
Correspondingly, there is no intermediate time plateau in the self-intermediate scattering function in 2D,  
Fig.~\ref{cage}d. The final decay of $F_s(k;t)$,  which in 3D is well described by a stretched exponential,
is replaced by a very slow decay in 2D. The intermediate time plateau in the mean-square displacement observed in
3D is replaced by an extended sub-diffusive regime in 2D, Fig.~\ref{cage}e. However, an intermediate time
plateau is observed in the correlation function quantifying bond-orientational 
correlations in 2D, $C_\Psi(t)$, Fig.~\ref{cage}f (see Methods for the definition of $C_\Psi$). 
Shown in Supplemental Figure 1a-c are $F_s(k;t)$ and $C_{\Psi}(t)$ for three additional 
2D glass formers and they behave similarly. 
Qualitatively different behavior of the translational and bond-orientational 
relaxation in 2D glass forming fluids is analogous to the absence of the
translational and the presence of the bond-orientational long-range order in 2D solids \cite{Strandburg1988}.

To quantify decoupling between translational and bond-orientational relaxation we compare the 
temperature dependence of the relaxation times characterizing $F_s(k;t)$ and $C_{(Q,\Psi)}(t)$,
where $Q$ and $\Psi$ refer to 3D and 2D 
correlation functions. 
We define the translational relaxation time $\tau_\alpha$ through the relation $F_s(k;\tau_\alpha) = e^{-1}$
and the bond-orientational relaxation 
time $\tau_\theta$ through $C_{(Q,\Psi)}(\tau_\theta) = e^{-1}$.
At the highest temperatures the ratio $\tau_\theta/\tau_\alpha$ is less than one for both 
the 3D and the 2D glass-former, Fig.~\ref{bond}a. However, this ratio stays approximately constant 
with decreasing temperature for the 3D glass-former, but grows monotonically for the 2D glass-former.
In Supplemental Figure 1d we show this ratio for the other 2D glass formers and show that the decoupling
is a general feature of 2D glassy dynamics.  
In addition, in Figs.~\ref{bond}b,c we show that the final translational and orientational relaxation satisfies 
the time-temperature superposition in 3D but not in 2D, and we show corresponding figures for
an additional glass former in 2D and 3D in Supplemental Figure 2. Fig.~\ref{bond}c clearly demonstrates the 
decoupling of the temperature dependence of the translational and bond-orientational relaxation times in 2D.

{\bf Dynamic heterogeneities.} The non-exponential decay of $F_s(k;t)$ is frequently attributed to the 
emergence of domains, referred to as dynamic heterogeneities, in which the relaxation is spatially
correlated and significantly different (faster or slower) than the average relaxation.   
While we find non-exponential decay in $F_s(k;t)$ for 3D and 2D
glass-formers, the nature of the decay is very different 
and this difference  
is mirrored by differences in the heterogeneous dynamics. 

\begin{figure}
\includegraphics[width=3.2in,bb=0.5in 2.0in 3.5in 7.4in,clip=True]{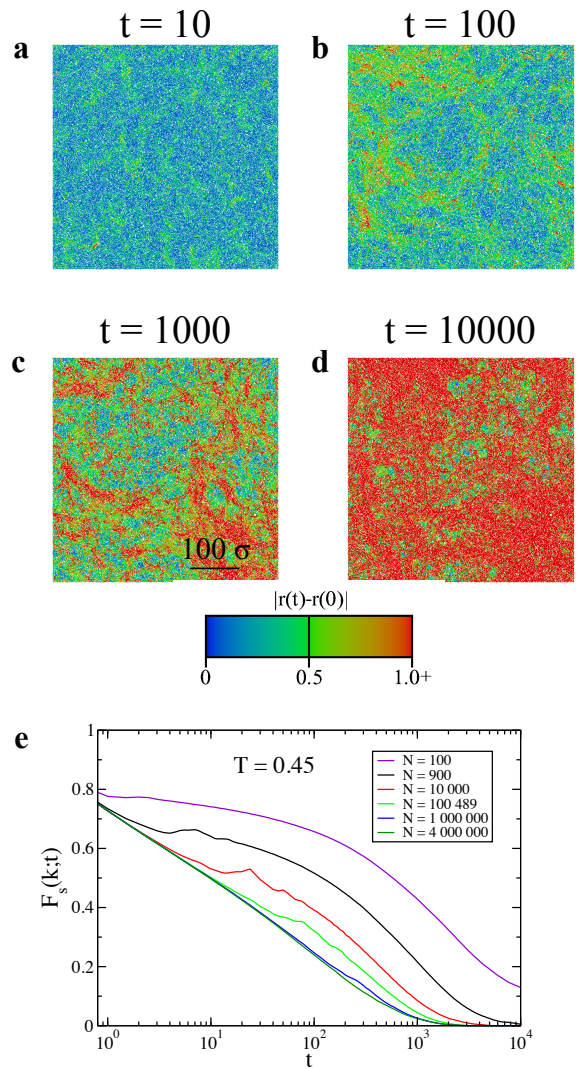}
\caption{\label{dh}\textbf{Dynamic heterogeneities.} {\bf a-d},
Displacement maps of the center of a system of 4 million particles 
that show the position of the particles at $t=0$
colored by the displacement of the particle at a later time $t$
for $T=0.45$. There are approximately $250\, 000$ particles in each
map. The scale
bar in \textbf{d} corresponds to 100 larger particle diameters. 
{\bf e}, The self-intermediate scattering function $F_s(k;t)$ calculated for systems of 100 to 4 million 
particles. There are clear finite size effects for less than one  million particles.}
\end{figure}

Shown in Figs.~\ref{dh}a-d are displacement maps showing the center of 
a four million particle simulation in 2D at $T=0.45$. The maps are created by 
coloring the particles, whose position is shown on $t=0$, according to the magnitude of 
their displacements
$|\mathbf{r}_n(t) - \mathbf{r}_n(0)|$ at a time $t$. The red particles have moved 
a distance equal to or greater than the diameter of a larger particle.
There are large domains of particles that have moved
less than a particle diameter even at $t = 10\, 000$.

Considering the large dynamically heterogeneous regions in Figs.~\ref{dh}a-d,
it is unsurprising that we also find large finite size effects. Shown in Fig.~\ref{dh}e
is $F_s(k;t)$ calculated for different size systems at the same temperature 
as shown in Figs.~\ref{dh}a-d. A plateau reminiscent of 
the plateau in 3D systems is present for the smaller systems but gradually disappears with increasing system size.
Similar finite size effects are also evident in the mean square displacement, Supplemental Figure 3a, and the
inherent structure dynamics, Supplemental Figure 3b.

\begin{figure*}
\includegraphics[width=7.5in,bb=0.3in 5.75in 7.0in 7.4in,clip=True]{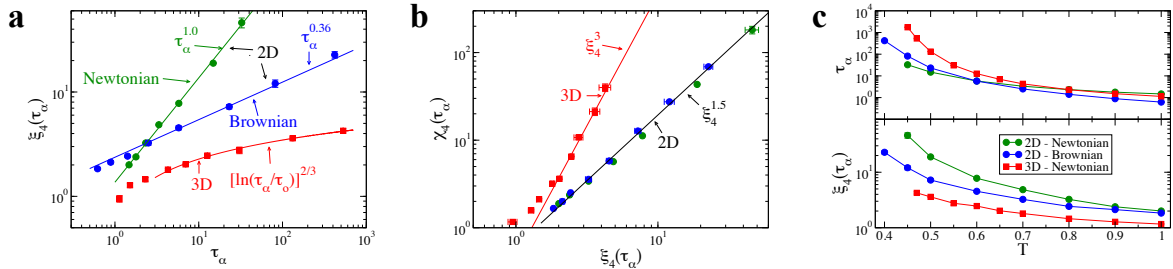}
\caption{\label{dh1}{\bf Properties of dynamic heterogeneities.}
{\bf a}, The dynamic correlation length $\xi_4(\tau_\alpha)$ versus the translational relaxation time $\tau_\alpha$.
The circles are the results for the 2D system where the underlying dynamics is Newtonian (green)
and Brownian (blue). The lines are fits to a power 
law $\xi_4(\tau_\alpha) = a \tau_\alpha^\beta$.
The red squares are results for the 3D system
where the underlying dynamics is Newtonian. 
The red line is a fit to $\xi_4(\tau_\alpha) = b [\ln(\tau_\alpha/\tau_o)]^{2/3}$.
{\bf b}, The dynamic susceptibility $\chi_4(\tau_\alpha)$ 
versus the dynamic correlation length $\xi_4(\tau_\alpha)$ 
for the 2D system where the underlying dynamics are
Newtonian (green circles) and Brownian (blue circles). 
The red squares are results for the 3D system. The lines
are fits to the power laws $\chi_4(\tau_\alpha) \sim \xi_4(\tau_\alpha)^3$
in 3D and $\chi_4(\tau_\alpha) \sim \xi_4(\tau_\alpha)^{1.5}$ in 2D.   
{\bf c}, The relaxation time $\tau_\alpha$ and the dynamic correlation 
length $\xi_4(\tau_\alpha)$ where the underlying dynamics are 
Newtonian (green) and Brownian (blue) for the 2D system, 
and Newtonian for the 3D system (red).}
\end{figure*}

To quantify dynamic heterogeneity shown in Fig.~\ref{dh}a-d
we use a four-point structure factor $S_4(q;t)$  \cite{Flenner2014} constructed from 
overlap functions $w_n(a;t) = \Theta(a-|\mathbf{r}_n(t) - \mathbf{r}_n(0)|)$,
where $\Theta(\cdot)$ is Heaviside's step function. 
The parameter $a$ is chosen such that 
$N^{-1} \left<\sum_n w_n(a;\tau_\alpha) \right> \approx F_s(k;\tau_\alpha)$, which results in $a=0.25$ in 3D 
and $a=0.22$ in 2D.  
To characterize the slow domains we calculate 
$S_4(q;t) = N^{-1} \left< \sum_n \sum_m w_n(a;t) w_m(a;t)
e^{i\mathbf{q} \cdot (\mathbf{r}_{n}(0)-\mathbf{r}_m(0))} \right>$ 
(note that $w_n(a;t)$ restricts the 
sums over the particles that moved less than $a$ over a time $t$).
The characteristic size of dynamically heterogeneous regions is quantified through the dynamic correlation length 
$\xi_4(t)$, which is determined from fitting $S_4(q;t)$ for small
$q$ to the Ornstein-Zernicke form $\chi_4(t)/\{1+[q \xi_4(t)]^2\}$. Here 
$\chi_4(t)$ is the dynamic susceptibility, which characterizes the overall strength of the dynamic
heterogeneity.
\begin{figure*}
\includegraphics[width=7.5in,bb=0.3in 5.75in 7.0in 7.4in,clip=True]{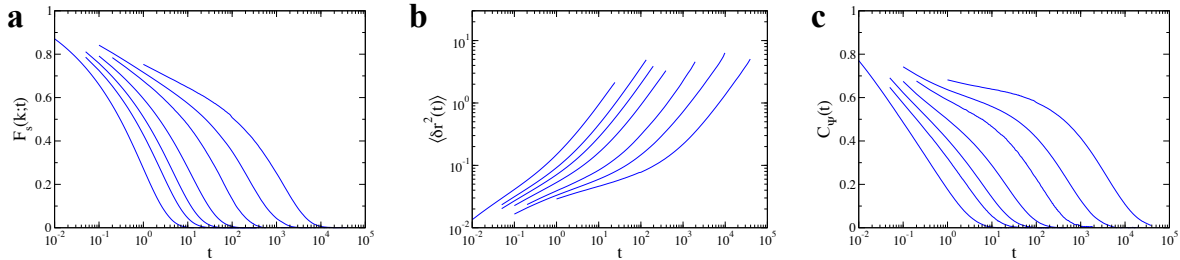}
\caption{\label{cage2dbd}\textbf{Structural relaxation in two dimensions with Brownian dynamics.} 
{\bf a}, The self-intermediate scattering function $F_s(k;t)$ 
for $T=1.0$, 0.8, 0.7, 0.6, 0.5, 0.45 and 0.4 listed from left to right.
{\bf b}, The mean square displacement $\left< \delta r^2(t) \right>$ showing the same temperatures as in (a).
{\bf c}, The bond angle time correlation function $C_\Psi(t)$ showing the same temperatures as in {\bf a}.
We emphasize that the 6 higher temperatures showed in this figure are the same as the temperatures showed 
in Figs.~\ref{cage}({\bf d})-({\bf f}). The finite size effects are much more pronounced in systems evolving with
Newtonian dynamics. This fact made impossible to simulate the 2D system at $T=0.4$ with Newtonian dynamics.
}
\end{figure*}

In Fig.~\ref{dh1}a we show the correlation between the translational
relaxation time, $\tau_\alpha$, and the dynamic correlation length calculated at 
$\tau_\alpha$, $\xi_4(\tau_\alpha)$,
for the 3D and 2D glass forming fluids. While for the 3D system we find that a power law is a poor description for
an extended range of $\tau_\alpha$, and a better description is 
$\xi_4(\tau_\alpha) \sim \left[\ln(\tau_\alpha/\tau_0) \right]^{2/3}$ (red line in Fig.~\ref{dh1}b),
we find that a power law $\xi_4(\tau_\alpha) \sim \tau_\alpha^\beta$ with $\beta=1.0\pm0.1$
describes the full range of results well for the 2D system. We show results for additional
glass formers in 2D and 3D in Supplemental Figure 4. 
Note that similar power law behavior was observed in simulations of 2D granular fluids \cite{Avila2014}.
In Fig.~\ref{dh1}b
we show that the relationship between the dynamic susceptibility and the dynamic correlation 
length is fundamentally different in 3D and 2D. For 3D systems 
$\chi_4(\tau_\alpha) \sim \xi_4(\tau_\alpha)^3$
at low temperatures, which implies compact dynamically heterogeneous regions. 
For 2D systems we observe $\chi_4(\tau_\alpha) \sim \xi_4(\tau_\alpha)^{1.5}$,
which suggests more ramified dynamically heterogeneous regions, see Fig.~\ref{dh}d.

{\bf Dependence on the microscopic dynamics.}
Lastly, we discuss the dependence of the long-time relaxation in 2D on the underlying microscopic dynamics
and two important consequences.
In 3D, an important
finding is that the long-time dynamics does
not depend on the microscopic dynamics; the same long-time dynamics has been observed in simulations 
using Newtonian \cite{Kob1994}, stochastic \cite{Gleim}, Brownian \cite{SFEPL} and Monte Carlo \cite{BerthierKob} 
dynamics. This result can be rationalized within the mode-coupling approach \cite{SL}. 
Surprisingly, we find that in 2D the long-time dynamics is quite different in the case of microscopic Newtonian 
and Brownian dynamics. The results corresponding to the those shown in Fig.~\ref{cage}d-f
for the Newtonian case are shown in Fig.~\ref{cage2dbd} for the Brownian case. 
Notably, the decay of $F_s(k;t)$ is strikingly different for the Brownian simulations 
than for the Newtonian simulations.

Importantly, the temperature dependence of the translational relaxation time 
is also decoupled from orientational relaxation time  
in the case of Brownian dynamics, Fig.~\ref{bond}a, but the 
ratio $\tau_\theta/\tau_\alpha$ is not as large for Brownian dynamics
than for Newtonian dynamics. 
In addition, we find a power law relationship $\xi_4(\tau_\alpha) \sim \tau_\alpha^\beta$ 
between the dynamic correlation length and the relaxation time but with $\beta=0.36\pm0.05$, which is a different
exponent than obtained for Newtonian dynamics, see Fig.~\ref{dh1}a. However, we find that the 
relationship between the strength of the dynamic heterogeneity and the dynamic correlation length in 2D
is the same for Brownian and Newtonian dynamics, see  Fig.~\ref{dh1}b. The latter two results show that
the universality of the relationships between the relaxation time and properties of heterogeneous dynamics
that we found in 3D \cite{Flenner2014} is absent in 2D.
Furthermore, a full description
of heterogeneous dynamics in 2D must also include the influence of the microscopic dynamics,
and descriptions solely in terms of the structure or the potential energy landscape are not sufficient in 2D. 

\section{Discussion}
Glassy dynamics in 
2D and in 3D are profoundly different. While we only presented detailed results for one glass-former, 
we verified that the features of the 
translational relaxation and dynamic heterogeneity are qualitatively the same 
for three additional 2D glass-formers (see Methods for their description and Supplemental Information for
the results). Our results call for a
re-examination of the present glass transition paradigm in 2D. We note that there is currently
no theoretical framework that accounts for the different dynamics observed in the 2D glass forming 
systems. However, we note that the dynamic picture of the Random First Order Transition theory breaks down 
for dimensions less than two, and has been described as marginal for two dimensions \cite{Kirkpatrick1987,Lubchenko2014}. 
Moreover, insights gained from theoretical analysis of the 2D glassy dynamics and glass
transition might shed light onto slow dynamics and the glass 
transition in 3D. It will also be interesting to investigate if the differences between 2D and 3D glassy dynamics  
are observable for glass-forming fluids in confinement and at interfaces or surfaces, \textit{i.e.} for
quasi-two-dimensional systems. 

We gratefully acknowledge the support of NSF grant CHE 1213401.
This research utilized the CSU ISTeC Cray HPC System supported by NSF Grant CNS-0923386.

\section{Methods}
\small
\subsection{Simulations}
We simulated binary mixtures of Lennard-Jones particle in two and three dimensions. 
The interaction potential is 
$V_{\alpha \beta}(r) = 4 \epsilon_{\alpha \beta}[(\sigma_{\alpha \beta}/r)^{12} - (\sigma_{\alpha \beta}/r)^6]$
where $\epsilon_{BB} = 0.5 \epsilon_{AA}$, $\epsilon_{AB} = 1.5 \epsilon_{AA}$, $\sigma_{BB} = 0.88 \sigma_{AA}$,
and $\sigma_{AB} = 0.8 \sigma_{AA}$. The results are presented in reduced units where $\sigma_{AA}\equiv\sigma$
is the unit of length and $\epsilon_{AA}$ the unit of energy. The unit of time for the Newtonian dynamics
simulations is $\sqrt{\sigma^2 m/\epsilon_{AA}}$ and the mass $m$ is the same for both species. 
The Newtonian dynamics simulations \cite{AT} were performed using LAMMPS \cite{Plimpton1995} 
for the 2D and 3D simulations and HOOMD-blue for the 2D simulations \cite{hoomd}. 
The LAMMPS simulations were run in an NVE ensemble, but there
is significant energy drift for the HOOMD-blue simulations for the lowest temperatures. Therefore, we ran the 
HOOMD-blue simulations using an NVT Nos\'e-Hoover thermostat with a coupling constant $\tau =10$. We 
ran at least one LAMMPS NVE simulation at every temperature to make sure that the conclusions did not
depend on the thermostat. All the results are averages over four or more production runs. 
The equations of motion for the Brownian dynamics simulations \cite{AT} are
$\dot{\mathbf{r}}_n(t) = \gamma^{-1} \mathbf{F}_n(t) + {\boldsymbol \eta}_n(t)$, 
where $\gamma = 1$ is the friction coefficient,
$\mathbf{F}_n(t)$ is the force on particle $n$ at time $t$, and ${\boldsymbol \eta}_n$ is 
a random noise term. The random noise satisfies the fluctuation dissipation relation 
$\left< {\boldsymbol \eta}_n(t) {\boldsymbol \eta}_m(t^\prime) \right> 
= 2 k_B T \gamma^{-1} \delta(t-t^\prime) \delta_{nm} \mathbf{1}$ where $\mathbf{1}$ is the unit tensor. 
The unit of time for the Brownian dynamics simulation is $\sigma^2 \gamma/\epsilon_{AA}$. 
The Brownian dynamics simulations were run using a modified version of LAMMPS and
our in house developed code. 

We simulated 2D systems of $10\, 000$ particles for $T \ge 0.9$ and
$250\, 000$ particles for $0.5 \le T \le 0.8$. At $T=0.45$ 
we studied 4 million particles for the Newtonian dynamics simulations, but
$250\, 000$ particles for the Brownian dynamics simulations.  We 
simulated $27\, 000$ particles in 3D using Newtonian dynamics. To check that the results
are independent of the system size for each state point for the 2D Newtonian dynamic simulations 
we ran $100\, 489$ particle simulations and checked to see if the results agreed with the
$250\, 000$ particle simulations. At $T=0.45$ they did not agree, and we increased the system size
until we found agreement between the 4 million particle system and an 8 million particle system.
For the 2D Brownian dynamics simulations we found agreement between $10\, 000$ particle simulations
and $250\, 000$ particle simulations for $T \ge 0.45$. For $T=0.4$ we found that a $100\, 489$ particle
simulation agreed with a $250\, 000$ particle system. 

We also examined the translational dynamics, bond-orientational, and dynamic heterogeneities for 
three additional systems in 2D and one additional glass forming
system in 3D. 
The first system is the one studied in Ref.~\cite{Candelier2010} and consists of a
32.167:67.833 binary mixture with the potential $V_{\alpha \beta}(r) = \epsilon (\sigma_{\alpha \beta}/r)^{12}$.
The size ratios are $\sigma_{AB} = 1.1 \sigma_{AA}$ 
and $\sigma_{BB} =  1.4 \sigma_{AA}$. We simulated this system using 
$250\, 000$ particles in 2D and $100\, 000$ particles 3D. 
The number density $\rho = 0.719 \sigma_{AA}^{-2}$
in 2D and $\rho = 0.719 \sigma_{AA}^{-3}$ in 3D. 
The second is a system introduced in Ref.~\cite{Harrowell1998},
which consists of an 50:50 mixture of repulsive particles where the potential 
$V_{\alpha \beta}(r) = \epsilon (\sigma_{\alpha \beta}/r)^{12}$. The size ratios are given by 
$\sigma_{AB} = 1.2 \sigma_{AA}$ and $\sigma_{BB} =  1.4 \sigma_{AA}$ and 
the number density $\rho = N/L^2  =0.74718 \sigma_{AA}^{-2}$. We simulated $250\, 000$ 
particles for this second additional system. 
The third system is the one introduced
in Ref.~\cite{Ohern}, which consists of an 50:50 mixture of harmonic spheres with
the interaction potential $V_{\alpha \beta}(r) = 0.5 \epsilon (1-r/\sigma_{\alpha \beta})^2$ 
for $r \le \sigma_{\alpha \beta}$ and $V_{\alpha \beta}(r) = 0$ otherwise. The size 
ratios are given by $\sigma_{AB} = 1.2 \sigma_{AA}$ and $\sigma_{BB} = 1.4 \sigma_{AA}$
and $\rho = 0.699 \sigma_{AA}^{-2}$.  We simulated $250\, 000$ particles 
for this third additional system. Some results for the three systems described in this paragraph
are given in the Supplemental Material. 

\subsection{Bond-Orientational Correlation Functions}
To measure bond-orientational relaxation times in 2D we first define 
$\Psi_6^n(t) = (N_b^n)^{-1} \sum_m e^{i6\theta_{nm}(t)}$, 
where $\theta_{nm}(t)$ is the angle between particle $n$ and particle $m$ at a time $t$, 
$N_b^n$ is the number of neighbors
of particle $n$, and the sum is over the neighbors of particle $n$ at the time $t$.
The neighbors are determined through Voronoi tessellation \cite{Bernard2011}.
The time dependence of the bond angle correlations was monitored by calculating 
$C_{\Psi}(t) = \left< \sum_n \Psi_6^n(t) [\Psi_6^n(0)]^* \right>/ \left<\sum_n |\Psi_6^n(0) |^2 \right>$
where $^*$ denotes the complex conjugate.

To measure bond-orientational relaxation in 3D we 
define $Q_{lm}^i(t) = (N_b^i)^{-1} \sum_j q_{lm}[\theta_{ij}(t),\phi_{ij}(t)]$ where $q_{lm}(\theta,\phi)$
are the spherical harmonics \cite{Steinhardt1983} 
and the sum is over the neighbors of a particle $i$ at a time $t$
determined through Voronoi tessellation. Next, we define the correlation function 
$\mathcal{Q}_l(t) = \left< [(4 \pi)/(2l +1)] \sum_i \sum_{m = -l}^{l} Q_{lm}^i(t) [Q_{lm}^i(0)]^* \right>$.
We calculated $C_Q(t) = \mathcal{Q}_6(t)/\mathcal{Q}_6(0)$ 
to monitor the decay of orientational correlations. 

We note that the conclusions remain 
unchanged if we define neighbors as being less than a distance equal to the first minimum of 
the pair correlation function rather than through Voronoi tessellation.

\newpage
\begin{figure*}
\includegraphics[width=1.\textwidth]{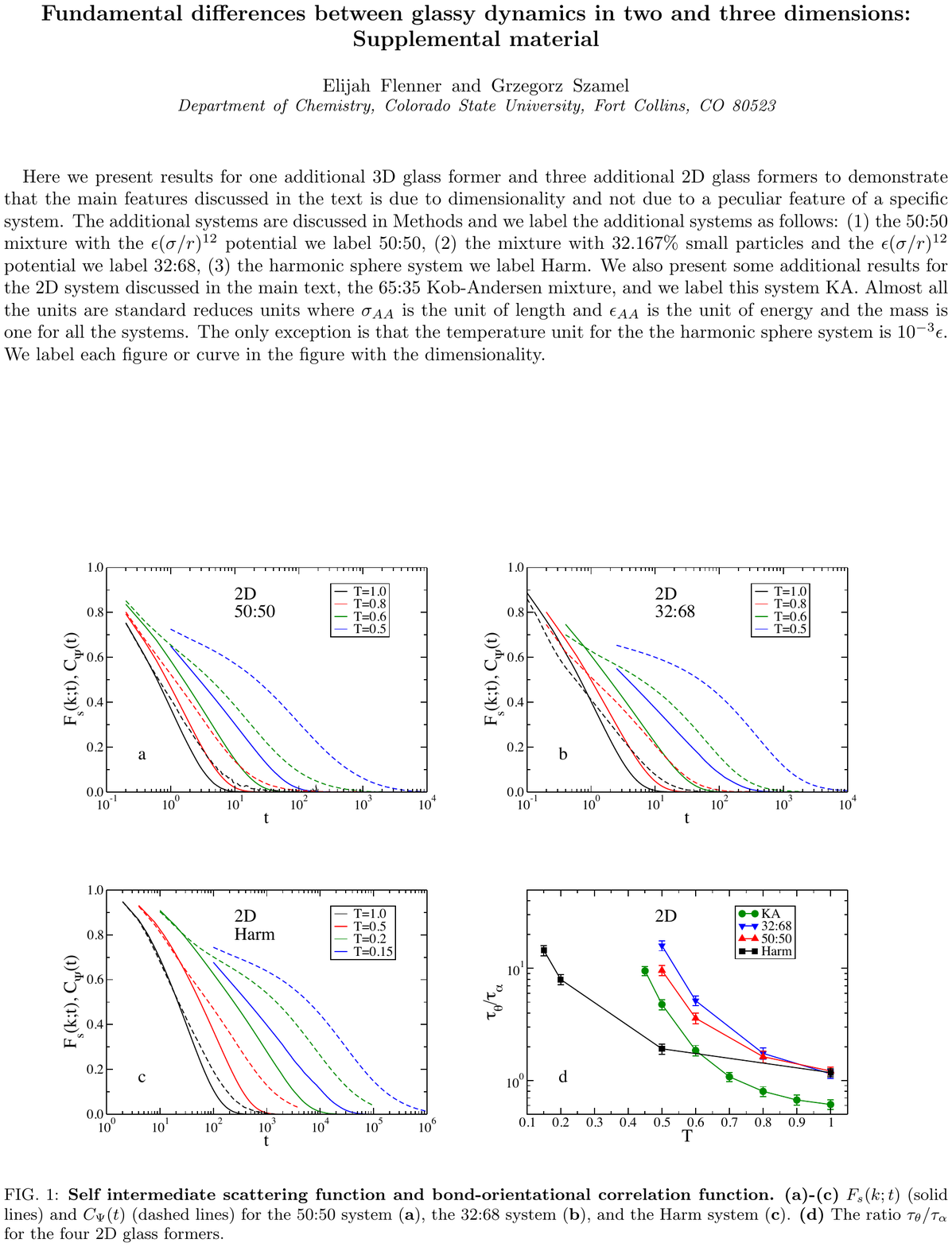}
\end{figure*}

\begin{figure*}
\includegraphics[width=1.\textwidth]{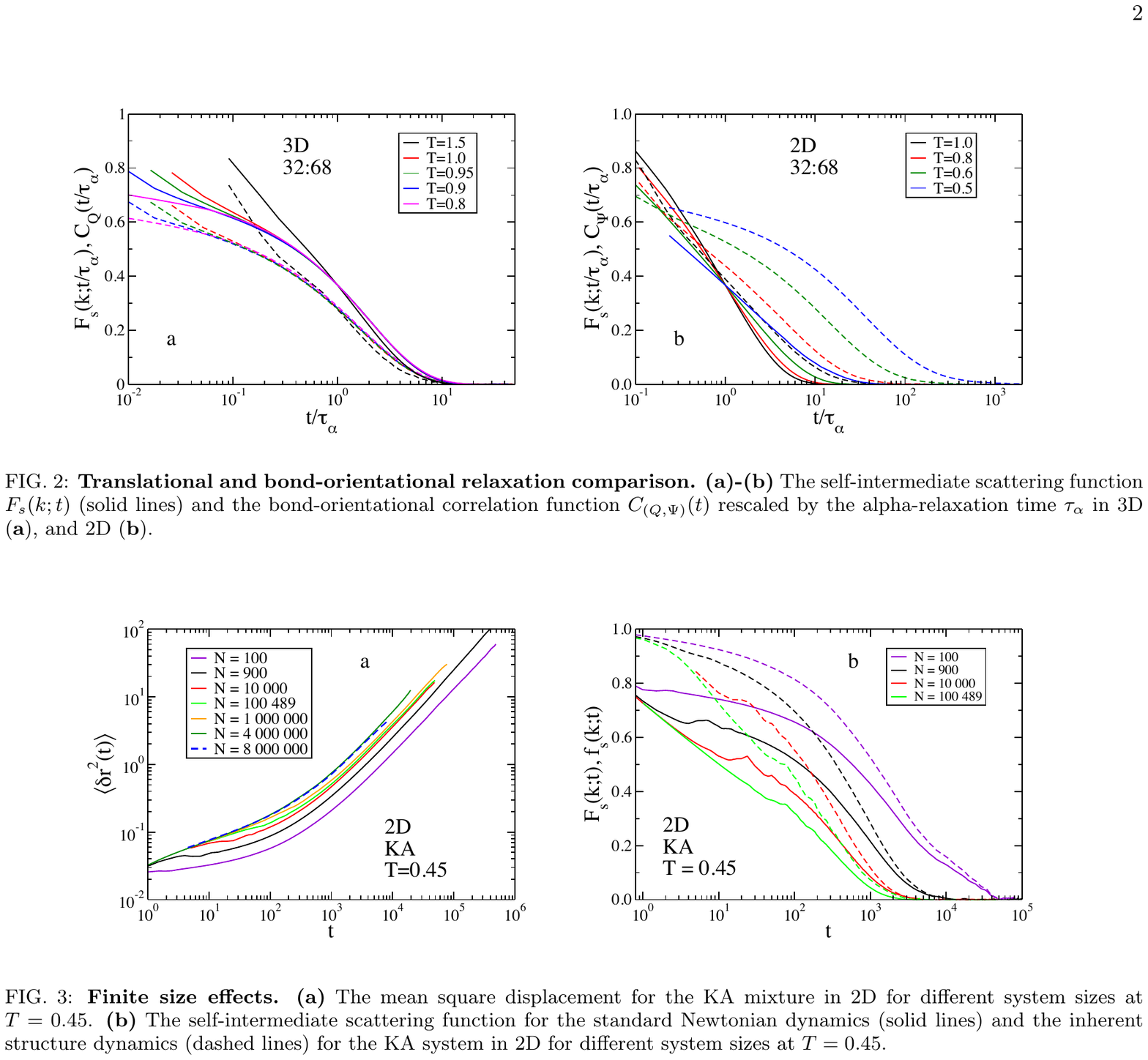}
\end{figure*}
\pagestyle{empty}

\begin{figure*}
\includegraphics[width=1.\textwidth]{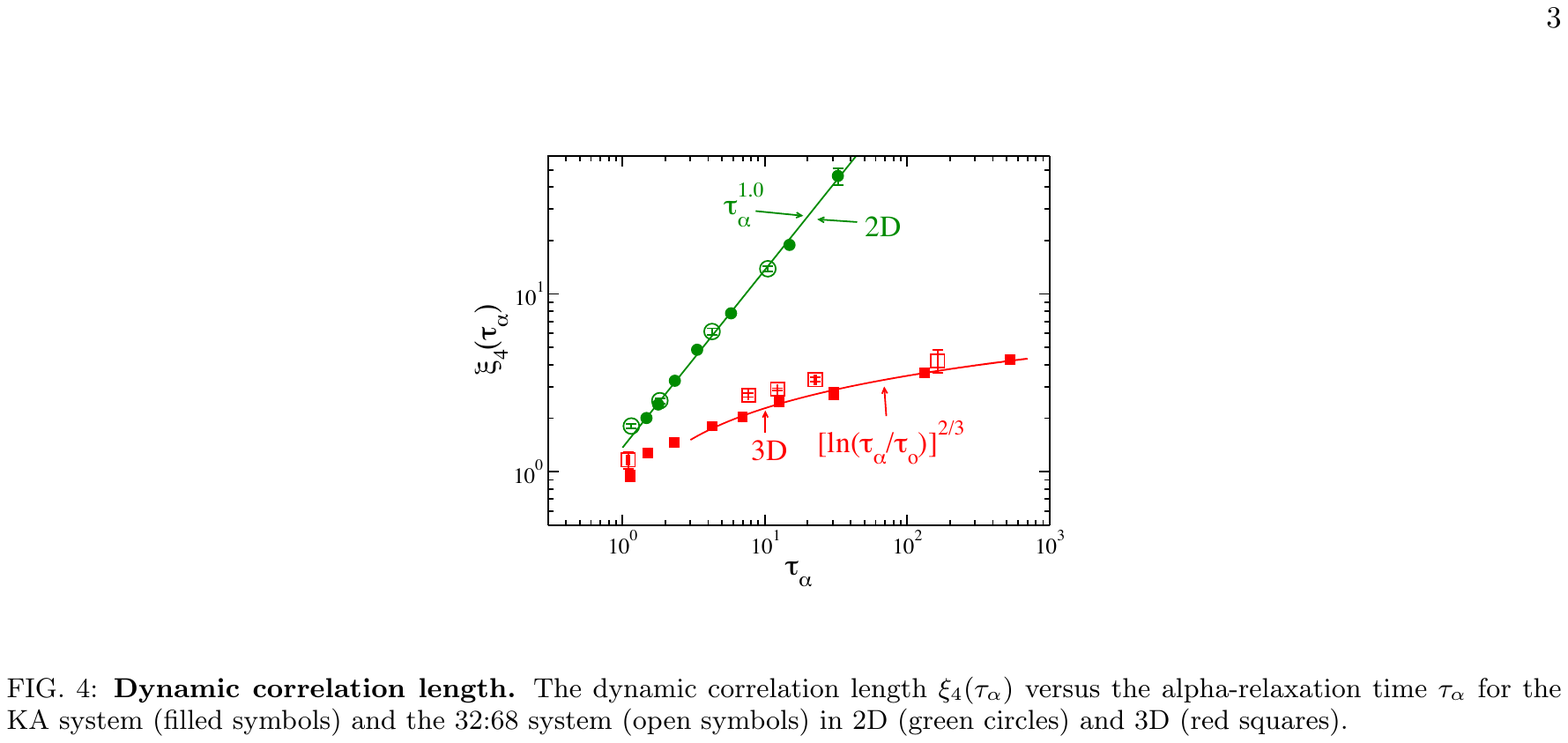}
\end{figure*}

\end{document}